\documentclass[nofootinbib,reprint,amsmath,amssymb,aps]{revtex4-1}
\usepackage{graphicx}
\usepackage[utf8,latin1]{inputenc}
\usepackage{dcolumn}
\usepackage{bm}
\usepackage{hyperref}
\usepackage{amsmath}
\newcommand{\qm}[1]{``#1''}
\usepackage{hyperref}
\hypersetup{colorlinks, linkcolor={red},citecolor={blue},urlcolor={blue}}

\begin{document}

\preprint{APS/123-QED}

\title[Poynting-Robertson effect as a dissipative system in general relativity]{Poynting-Robertson effect as a dissipative system in general relativity}

\author{Vittorio De Falco$^{1}$}\email{vittorio.defalco@physics.cz}
\author{Emmanuele Battista$^{1,2,3}$\vspace{0.5cm}}\email{emmanuelebattista@gmail.com}

\affiliation{$^1$ Research Centre for Computational Physics and Data Processing, Faculty of Philosophy \& Science, Silesian University in Opava, Bezru\v{c}ovo n\'am.~13, CZ-746\,01 Opava, Czech Republic\\
$^2$ Universit\`{a} degli studi di Napoli \qm{Federico II}, Dipartimento di Fisica \qm{Ettore Pancini}, Complesso Universitario di Monte S. Angelo, Via Cintia Edificio 6, 80126 Napoli, Italy\\
$^3$ Istituto Nazionale di Fisica Nucleare, Sezione di Napoli, Complesso Universitario di Monte S. Angelo, Via Cintia Edificio 6, 80126 Napoli, Italy
}

\date{\today}

\begin{abstract}
We determine for the first time in the literature the analytic form of the Rayleigh potential of the general relativistic Poynting-Robertson effect. The employed procedure is based on the use of an integrating factor and a new integration strategy where the test particle's dissipated energy represents the fundamental variable. The obtained results and their implications are discussed. Finally, concluding remarks and future projects are drawn.

\end{abstract}

\maketitle

\section{Introduction}
\label{sec:intro}
Dissipation is a subject which concerns several research fields, ranging from classical to quantum physics. Usually, it deals with the waste of stored (mechanical) energy undergone by a dynamical system during time evolution, but its meaning can be also understood in a broader sense, depending on the context and the problem to be investigated. However, dissipation configures as a fundamental ingredient to make a model more realistic, although the mathematical framework spreads more and more out of control, showing some critical consequences, like: loss of existence, smoothness, and symmetries of the original solution, presence of topologically complex structures featuring chaotic behaviour, difficulties in the numerical integration process \cite{Evans2010,Sprott2011,Quarteroni2009}. In particular, in General Relativity (GR) dissipative forces strongly couple with the geometrical structure of the background spacetime, giving rise to highly non-linear functions. 

In this paper, we consider the following general problem: given the equations of motion of a dissipative system in GR, we would like to derive them from the principle of least action through the Euler-Lagrange equations
\begin{equation} \label{eq:ELequation}
\frac{d}{dt}\left(\frac{\partial \mathcal{L}}{\partial \dot{q}^h}\right)-\frac{\partial \mathcal{L}}{\partial q^h}=-\frac{\partial \mathcal{F}}{\partial \dot{q}^h},\quad h=1,\dots,N,
\end{equation}
where $N$ denotes the  degrees of freedom of the system and the unknown functions are: the Lagrangian $\mathcal{L}(\boldsymbol{q},\boldsymbol{\dot{q}})$ (encompassing the kinetic energy and the conservative/generalized forces) and the Rayleigh dissipation function $\mathcal{F}(\boldsymbol{q},\boldsymbol{\dot{q}})$ (describing the dissipative/non-conservative forces) \cite{Goldstein2002}. This is the renowned inverse problem of the calculus of variations \cite{Santilli1978,Morandi1990,Do2016}, which can be proven to be a \emph{well-posed problem} in the sense of Hadamard (see Refs. \cite{Gitman2007b,Mestdag2011,Gitman2007,Kochan2010,Kabanikhin2011} and references therein). 

The actual increasing computational power and advanced numerical methods entail a widespread attitude to approach such kinds of difficult problems by entrusting them mainly through appropriate numerical codes. Despite being a precious resource, these sometimes discourage theoretical investigations. A natural consequence is that analytic results, which are fundamental to have direct insight into mathematical and physical details of the model under investigation, are becoming more and more rare. It is in that spirit that we have determined for the first time in the GR literature the analytical form of the Rayleigh potential related to the general relativistic Poynting-Robertson (PR) effect.

This phenomenon deals with the motion of relatively small-sized test particles (e.g., dust grains or gas clouds \cite{Liou1995,Kimura2002,Klavoka2013}, meteors \cite{Wyatt1950,Yajima2014}, accretion disk matter elements \cite{Rafikov2011,Lancova2017}) around radiating massive sources. 
The radiation field, which is directed radially outward the source, besides exerting a force against the gravitational pull generates also a drag force opposite to the test particle orbital motion (triggered by the process of  absorption and re-emission of the incoming radiation) \cite{Poynting1903,Robertson1937}. The PR effect removes very efficiently angular momentum and energy from the test particle, configuring thus as a dissipative force. The general relativistic PR effect models have been proposed from the two-dimensional \cite{Bini2009,Bini2011} to the three-dimensional cases in Kerr spacetime \cite{DeFalco20183D,Bakala2019,Wielgus2019}.

Recently, a Lagrangian formulation of the PR effect has been provided \cite{Defalco2018}. The novel aspects of such approach consists in the introduction of an integrating factor. Furthermore, in a separate letter \cite{DBletter2019} we have presented a brief and basic account of our new strategy to determine the Rayleigh dissipation function. Here, we aim at giving a more detailed derivation and comprehensive analysis of the obtained results.

The article is organised as follows: in Sec. \ref{sec:AEF} we extensively explain how to derive the Rayleigh potential of the general relativistic PR effect; in Sec. \ref{sec:int_res} we discuss and interpret the obtained results; in Sec. \ref{sec:end} we finally draw our conclusions and comment on future projects. 

\section{Derivation of the general relativistic PR effect's Rayleigh potential}
\label{sec:AEF}
In this section, we establish the main result of the paper, i.e., the analytical form of the Rayleigh potential for the general relativistic PR effect (see Eq. (\ref{eq:Rayleigh_potential_PR}) below). First of all, we briefly explain the general relativistic PR effect model (Sec. \ref{sec:GRPReffect}), afterwards we remind the reader how to determine the integrating factor (Sec. \ref{sec:expo}), and then in Sec. \ref{sec:GRRPO_sol} we explain our new strategy to derive the Rayleigh potential.  

\subsection{General relativistic PR effect model}
\label{sec:GRPReffect}
The general relativistic PR effect describes the dynamics of a test particle moving with a timelike velocity $\boldsymbol{U}$ around a rotating (or a static) compact object under the influence of a gravitational field, described by the Kerr metric (or the Schwarzschild metric)\footnote{Here, we consider metrics with signature +2, therefore in our notations a timelike vector $v^\alpha$ has norm $v^\alpha v_\alpha=-1$. We use also geometrical units $G=c=1$, for the gravitational constant $G$ and the speed of light in the vacuum $c$.}in a coordinates system $\boldsymbol{X}$, the radiation pressure, and the radiation drag force. The test particle equations of motion are $a(\boldsymbol{X},\boldsymbol{U})^\alpha=F_{\rm (rad)}(\boldsymbol{X},\boldsymbol{U})^\alpha$, where $a(\boldsymbol{X},\boldsymbol{U})^\alpha$ is the test particle acceleration and $F_{\rm (rad)}(\boldsymbol{X},\boldsymbol{U})^\alpha$ is the radiation force per unit mass, including the radiation pressure and the PR effect. Following the same line of reasoning of J. H. Poynting and H.P.  Robertson \cite{Poynting1903,Robertson1937}, we write the equations of motion first in the test particle rest frame and then in the static observer frame located at infinity. To this aim, we exploit the \emph{relativity of observer splitting formalism}, which represents a powerful method in GR to distinguish the gravitational effects from the fictitious forces arising from the relative motion of two non-inertial observers \cite{Jantzen1992,Bini1997a,Bini1997b,Defalco2018}. Such formalism allows us to derive the test particle equations of motion in the reference frame of the static observer located at infinity as a set of coupled first order differential equations \cite{Bini2009,Bini2011,DeFalco20183D,Bakala2019}. The radiation force is modelled as a pure electromagnetic field, where the photons move along null geodesics on the background spacetime. The stress-energy tensor reads as \cite{Bini2009,Bini2011,Defalco2018,DeFalco20183D,Bakala2019}
\begin{equation} \label{eq:set}
T^{\alpha\beta}=\Phi^2 k^\alpha k^\beta,
\end{equation}
where $k^\alpha$, which is a function of the local coordinates $\boldsymbol{X}$ only, denotes the photon four-momentum satisfying the conditions $k_\alpha k^\alpha=0$ and $k^\beta \nabla_\beta k^\alpha=0$, whereas  $\Phi$ represents a parameter related to the radiation field intensity. Therefore, the radiation force $F_{\rm (rad)}(\boldsymbol{X},\boldsymbol{U})^\alpha$ is given by
\begin{equation} \label{eq:radforce1}
\begin{aligned}
F_{\rm (rad)}(\boldsymbol{X},\boldsymbol{U})^\alpha&\equiv-\tilde{\sigma}\mathcal{P}(\boldsymbol{U})^\alpha_{\;\beta} T^\beta_{\;\nu} U^\nu\\
&=-\tilde{\sigma}\Phi^2\left(k^\alpha k_\nu U^\nu+U^\alpha U_\beta k^\beta k_\nu U^\nu\right),
\end{aligned}
\end{equation}
where $\mathcal{P}(\boldsymbol{U})^\alpha_{\; \beta}=\delta^\alpha_{\; \beta} +U^\alpha U_\beta$ is the projection operator on the spatial hypersurface orthogonal to $\boldsymbol{U}$, $\tilde{\sigma}=\sigma/m$ with $\sigma$ the Thomson scattering cross section describing the radiation field-test particle interaction and $m$ the test particle mass. Since the factor $-\tilde{\sigma}\Phi^2$ is a constant with respect to the test particle velocity field $\boldsymbol{U}$, we can ease the notations by considering only
\begin{equation} \label{eq:radforce2}
\tilde{F}_{\rm (rad)}(\boldsymbol{X},\boldsymbol{U})^\alpha\equiv k^\alpha k_\nu U^\nu+U^\alpha U_\beta k^\beta k_\nu U^\nu.
\end{equation}

The strength of the radiation force is characterised by the luminosity parameter $A$, defined as $A/M=L/L_{\rm EDD}\in[0,1]$, where $M$ is the mass of the central compact object, $L$ is the emitted luminosity measured by an observer at infinity, and $L_{\rm EDD}$ is the Eddington luminosity \cite{Bini2009,Bini2011,DeFalco20183D,Bakala2019}. The photons of the radiation field are emitted with an impact parameter $b$, which can assume the following values: either $b=0$ (radial case) \cite{Bini2009,DeFalco20183D} or $b\neq0$ (general case) \cite{Bini2011,Bakala2019}.

\subsection{Existence of general relativistic Rayleigh potential}
\label{sec:expo}
In this section, we describe some further developments of the framework illustrated in Ref. \cite{Defalco2018}. We show the procedure followed to calculate the integrating factor (Sec. \ref{sec:intfact}), and its expression in the classical limit (Sec. \ref{sec:WFA}), in order to recover its full form and interpret it physically.

\subsubsection{Integrating factor}
\label{sec:intfact}
The radiation force $\tilde{F}_{\rm (rad)}(\boldsymbol{X},\boldsymbol{U})^\alpha$ depends non-linearly on the test particle velocity field $\boldsymbol{U}$, therefore we check whether it can be expressed in terms of the Rayleigh potential $V$, i.e., $\tilde{F}_{\rm (rad)}(\boldsymbol{X},\boldsymbol{U})^\alpha=-\partial V/\partial U_\alpha$ \cite{Goldstein2002}. It is important to note that the components of (\ref{eq:radforce2}) can be seen as the components of a differential semi-basic one-form\footnote{A differential semi-basic one-form (also known in the literature as \emph{one-form along the tangent bundle projection} \cite{Martinez1992,MARTINEZ19931,Mestdag2011}) is defined through the map $\tilde{\boldsymbol{\omega}}:T\mathcal{M}\rightarrow T^*\mathcal{M}$, where $T\mathcal{M}$ and $T^*\mathcal{M}$ denote the tangent bundle and the cotangent bundle of the base spacetime manifold $\mathcal{M}$, respectively.} $\tilde{\boldsymbol{\omega}}(\boldsymbol{X},\boldsymbol{U})=\tilde{F}_{\rm (rad)}(\boldsymbol{X},\boldsymbol{U})^\alpha \boldsymbol{{\rm d}}X_\alpha$, 
which is defined over the simply connected domain $T\mathcal{M}$. Indeed, the base spacetime manifold $\mathcal{M}$ is represented by the whole space outside the compact object, including the event horizon, times the time line, whereas all the fibers $T_p\mathcal{M}$ in $p\in\mathcal{M}$ are $n$-dimensional hypercubes, since the limit velocity coincides with the speed of light. 

Since the cross derivatives of $\tilde{F}_{\rm (rad)}(\boldsymbol{X},\boldsymbol{U})^\alpha$ are not equal, i.e., $\partial \tilde{F}_{\rm (rad)}(\boldsymbol{X},\boldsymbol{U})^\alpha/\partial U_\lambda\neq \partial \tilde{F}_{\rm (rad)}(\boldsymbol{X},\boldsymbol{U})^\lambda/\partial U_\alpha$, the semi-basic one-form turns out to be not exact \cite{Defalco2018}. However, the introduction of an integrating factor $\mu$ could make the differential semi-basic one-form closed in its domain of definition, ensuring thus that it is exact. The components of the (exact) semi-basic one-form will be now represented by $\mu \tilde{F}_{\rm (rad)}(\boldsymbol{X},\boldsymbol{U})^\alpha$, i.e., we have
\begin{equation}
\boldsymbol{\omega}(\boldsymbol{X},\boldsymbol{U})\equiv\mu \tilde{\boldsymbol{\omega}}(\boldsymbol{X},\boldsymbol{U})=\mu \tilde{F}_{\rm (rad)}(\boldsymbol{X},\boldsymbol{U})^\alpha \boldsymbol{{\rm d}}X_\alpha,    
\label{PR_semibasic_oneform}
\end{equation}
and the condition according to which this \qm{upgraded differential semi-basic one-form} is closed yields
\begin{equation} \label{eq:difeqformu}
\begin{aligned}
0&=\left(-k^\alpha\frac{\partial\mu}{\partial U_\lambda}+k^\lambda\frac{\partial\mu}{\partial U_\alpha}\right)\\
&+U^\alpha\left(\frac{\partial \mu}{\partial U_\lambda}k^\beta U_\beta+2\mu k^\lambda\right)\\
&-U^\lambda\left(\frac{\partial \mu}{\partial U_\alpha}k^\beta U_\beta+2\mu k^\alpha\right).\\
\end{aligned}
\end{equation}
This in turn implies that $\mu$ should solve simultaneously the following two differential equations:
\begin{eqnarray} 
&&-k^\alpha\frac{\partial\mu}{\partial U_\lambda}+k^\lambda\frac{\partial\mu}{\partial U_\alpha}=0, \label{eq:difeqformu1}\\
&&\frac{\partial \mu}{\partial U_\lambda}k^\beta U_\beta+2\mu k_\lambda=0. \label{eq:difeqformu2}
\end{eqnarray}
The radiation force (\ref{eq:radforce2}) can be split into two parts
\begin{equation} \label{eq:splitforce}
\tilde{F}_{\rm (rad)}(\boldsymbol{X},\boldsymbol{U})^\alpha=\mathbb{F}_{\rm C}(\boldsymbol{X},\boldsymbol{U})^\alpha+\mathbb{F}_{\rm NC}(\boldsymbol{X},\boldsymbol{U})^\alpha,
\end{equation}
where
\begin{eqnarray}
\mathbb{F}_{\rm C}(\boldsymbol{X},\boldsymbol{U})^\alpha&\equiv& T^\alpha_{\;\nu} U^\nu=-k^\alpha \mathbb{E}(\boldsymbol{X},\boldsymbol{U}), \label{eq:twoforces1}\\
&&\notag\\
\mathbb{F}_{\rm NC}(\boldsymbol{X},\boldsymbol{U})^\alpha&\equiv& U^\alpha U_\beta T^\beta_{\;\nu} U^\nu=\mathbb{E}(\boldsymbol{X},\boldsymbol{U})^2 U^\alpha,\label{eq:twoforces2}
\end{eqnarray}
with 
\begin{equation} \label{particle_energy_1}
\mathbb{E}(\boldsymbol{X},\boldsymbol{U})\equiv \mathbb{E}=-k_\beta U^\beta, 
\end{equation}
representing the test particle energy \cite{Bini2009,Bini2011,Defalco2018} (more details regarding $\mathbb{E}$ will be given in Sec. \ref{sec:LOGPOT}). We refer to $\mathbb{F}_{\rm C}(\boldsymbol{X},\boldsymbol{U})^\alpha$ as the ``conservative'' part of the radiation force, due to its property to admit a primitive function without considering an integrating factor, since it depends linearly on $U^\alpha$; whereas $\mathbb{F}_{\rm NC}(\boldsymbol{X},\boldsymbol{U})^\alpha$ stands for the ``non-conservative'' components of the radiation force, because it is nonlinear and the related primitive can be determined only through the introduction of the integrating factor $\mu$. 

At this stage, we would like to determine a common integrating factor for both the components of the radiation force. However, it is noteworthy to stress that this request is not trivial at all. In fact, in principle we might come up with two different integrating factors $\mu_1$ and $\mu_2$, where $\mu_1$ is related to $\mathbb{F}_{\rm C}(\boldsymbol{X},\boldsymbol{U})^\alpha$ and solves Eq. (\ref{eq:difeqformu1}), while $\mu_2$ is associated with $\mathbb{F}_{\rm NC}(\boldsymbol{X},\boldsymbol{U})^\alpha$ and solves Eq. (\ref{eq:difeqformu2}). Therefore, the system of differential equations (\ref{eq:difeqformu1}) and (\ref{eq:difeqformu2}) admits in general two distinct solutions\footnote{In the most general case it may even happen that some of the differential equations, defining the various integrating factors, might not admit any solution at all.} ($\mu_1\neq\mu_2$). For instance, for the conservative components $\mathbb{F}_{\rm C}(\boldsymbol{X},\boldsymbol{U})^\alpha$ the function $\mu=\mbox{const}$ clearly represents a solution of (\ref{eq:difeqformu1}). Therefore, the possibility to find a unique solution, different from the trivial one $\mu=\mbox{const}$, for both $\mathbb{F}_{\rm NC}(\boldsymbol{X},\boldsymbol{U})^\alpha$ and $\mathbb{F}_{\rm C}(\boldsymbol{X},\boldsymbol{U})^\alpha$ is not so obvious \emph{a priori}. However, as it can be easily checked from Eqs. (\ref{eq:difeqformu1}) and (\ref{eq:difeqformu2}), the PR effect exhibits the peculiar propriety to have one common integrating factor for the two components (i.e., $\mu_1=\mu_2 \equiv \mu$), which, up to a constant, reads as\footnote{In Eq. (\ref{eq:if}) we have corrected a little error occurred in Ref. \cite{Defalco2018}.}
\begin{equation} \label{eq:if}
\mu =\frac{1}{\mathbb{E}^2}.
\end{equation}
The use of an integrator factor permits to guarantee, in a non-intuitive manner, existence and uniqueness (up to a constant term) of the Rayleigh potential. 

\subsubsection{Classical limit of the integrating factor}
\label{sec:WFA}
As we have just pointed out, the integrating factor (\ref{eq:if}) is defined up to a constant (with respect to the velocity field $\boldsymbol{U}$) which can be determined in the \emph{classical limit} (weak field approximation, $M/r\rightarrow0$, and non-relativistic velocities, $\nu/c\rightarrow0$, \cite{Defalco2018}). By employing the Schwarzschild metric in the equatorial plane $\theta=\pi/2$,
\begin{equation} \label{Schwarzschild_metric_1}
g_{\mu \nu}=\rm{diag}\left[-\left(1-\frac{2M}{r}\right),\left(1-\frac{2M}{r}\right)^{-1},r^2,r^2\right],
\end{equation}
the test particle and the photon velocities read, respectively, as \cite{Bini2009,Defalco2018,Chandrasekhar1992}\footnote{Due to the spherical symmetry of the metric, we are allowed, without loss of generality, to reduce the problem to a two dimensional setting so that all calculations are easily performed.}
\begin{eqnarray} 
U^\alpha&=&\left[\frac{\gamma}{\sqrt{1-\frac{2M}{r}}},\frac{\gamma\nu\sin\alpha}{\left(\sqrt{1-\frac{2M}{r}}\right)^{-1}},0,\frac{\gamma\nu\cos\alpha}{r}\right], \label{eq:vel-phot1}\\
k_\alpha&=&E_{\rm p}\left[-1,\frac{1}{1-\frac{2M}{r}},0,0\right], \label{eq:vel-phot2}
\end{eqnarray}
where $\gamma$ is the Lorentz factor, $\alpha$ and $\nu$ represent, respectively, the azimuthal angle measured clockwise from the positive $\boldsymbol{\hat{\varphi}}$ direction and the module of the test particle velocity in the spatial hypersurface $\boldsymbol{\hat{r}}-\boldsymbol{\hat{\varphi}}$ orthogonal to the zero angular momentum observers (ZAMOs), $E_{\rm p}=-k_t$ is the photon energy. 
Note that, without loss of generality, we have considered a radial radiation photon impact parameter (see \cite{Bini2009,Bini2011,Defalco2018}, for further details). 

Bearing in mind Eqs. (\ref{particle_energy_1}), (\ref{Schwarzschild_metric_1}), (\ref{eq:vel-phot1}), and (\ref{eq:vel-phot2}), it is easy to show that in the classical limit
\begin{equation} \label{eq:Elimit}
\mathbb{E}\approx E_{\rm p}\left(1-\dot{r}\right),
\end{equation}
where we have decided to neglect, from now on, all terms containing the factor $M/r$, since classically they give a higher-order contribution to the radiation force. We note that the classical Rayleigh potential can be easily recovered without the introduction of an integrating factor (see Ref. \cite{Defalco2018}), hence in this limit $\mu=1$. Since $\mu\sim\mbox{const}/(\mathbb{E}^2)$, this leads immediately to choose the constant term equals to $E_{\rm p}^2$, so that Eq. (\ref{eq:if}) can now be recasted as
\begin{equation} \label{eq:if2}
\mu =\frac{E_{\rm p}^2}{\mathbb{E}^2}.
\end{equation}
The appearance of a constant term having the physical dimension of the square of an energy could also be expected on general grounds, since we require a dimensionless integrating factor. 

\subsection{General relativistic Rayleigh potential}
\label{sec:GRRPO_sol}
In this section we determine the analytic form of the general relativistic Rayleigh potential of the PR effect. Typical examples of Rayleigh functions discussed in the literature involve simple models where the dynamical equations can be easily integrated yielding a polynomial function in the velocities \cite{Goldstein2002}. However, as pointed out before, the radiation force (\ref{eq:radforce2}) involves a nonlinear function of the test particle velocity field $\boldsymbol{U}$. Therefore, we have set up an original approach where the key variable is the energy dissipated by the system. Indeed, we shall see how the calculations regarding the determination of the $V$ potential related to the (exact) differential semi-basic one-form (\ref{PR_semibasic_oneform}) are greatly simplified by expressing the radiation force in terms of the energy variable (\ref{particle_energy_1}). This represents one of the crucial aspects of our procedure, because it reduces tremendously the calculations, passing from an integration involving the four variables $\boldsymbol{U}$ to only one, represented by the energy $\mathbb{E}$. 
  
\subsubsection{Preliminary considerations}
\label{sec:preliminary}

As pointed out in Sec. \ref{sec:expo}, the introduction of the integrating factor (\ref{eq:if}) (or equivalently (\ref{eq:if2})) makes the semi-basic one-form (\ref{PR_semibasic_oneform}) closed and hence, due to the topological properties of its domain $T\mathcal{M}$, exact. Therefore, it makes sense the research of a potential function $V(\boldsymbol{X},\boldsymbol{U})$ (i.e., the Rayleigh potential) such that, 
\begin{equation}
    - \dfrac{\partial V(\boldsymbol{X},\boldsymbol{U})}{\partial U_\alpha} = \mu \tilde{F}_{\rm (rad)}(\boldsymbol{X},\boldsymbol{U})^\alpha.
    \label{eq:potential_def}
\end{equation}
The last equation can be obtained in an elegant way by resorting to the tools of differential geometry \cite{Abraham1978,MARTINEZ19931,Mestdag2011}. Indeed, the closure condition (\ref{eq:difeqformu}) can be formulated concisely as
\begin{equation}
    \boldsymbol{{\rm d^V}}\boldsymbol{\omega} = 0,
\end{equation}
where the operator $\boldsymbol{{\rm d^V}}$ denotes the \emph{vertical exterior derivative}, whose local expression on a generic (smooth) 0-form (i.e., a function) $\mathcal{G}(\boldsymbol{X},\boldsymbol{U})$ defined on $T\mathcal{M}$ is given by
\begin{equation} \label{eq:vertical_derivative1}
\boldsymbol{{\rm d^V}}\mathcal{G}(\boldsymbol{X},\boldsymbol{U})= \frac{\partial \mathcal{G}(\boldsymbol{X},\boldsymbol{U})}{\partial U_\alpha} \, \boldsymbol{{\rm d}}X_\alpha.
\end{equation}

The Poincar\'e lemma, adapted to the case of vertical differentiation \citep{Martinez1992}, guarantees that $\boldsymbol{\omega}$ is also exact, i.e., it can be expressed as the vertical exterior derivative of a 0-form $V(\boldsymbol{X},\boldsymbol{U}$) (i.e., the primitive or potential function), namely
\begin{equation}
-\boldsymbol{{\rm d^V}}V(\boldsymbol{X},\boldsymbol{U})= \boldsymbol{\omega}(\boldsymbol{X},\boldsymbol{U}).
\label{eq:exact_cond}
\end{equation}
It is easy to see that (\ref{eq:potential_def}) stems from Eqs. (\ref{eq:vertical_derivative1}) and (\ref{eq:exact_cond}). 

As we said before, the main feature of our technique is represented by the fact that the fundamental variable embodying the dynamical aspects of the investigated system is the dissipated energy (\ref{particle_energy_1}). Therefore, we first need to express the components of the differential semi-basic one-form (\ref{PR_semibasic_oneform}) in terms of the energy $\mathbb{E}$. These formulae were given before in Eqs. (\ref{eq:twoforces1}) and (\ref{eq:twoforces2}). Furthermore, we consider the derivative operator in terms of the energy variable through the usual chain rule, i.e.,
\begin{equation} \label{eq:der}
\frac{\partial(\cdot)}{\partial U_\alpha}=\frac{\partial \mathbb{E}}{\partial U_\alpha} \frac{\partial(\cdot)}{\partial \mathbb{E}}=-k^\alpha \frac{\partial (\cdot)}{\partial  \mathbb{E}}.
\end{equation}  
The above equation permits to write (\ref{eq:potential_def}) as
\begin{equation} \label{eq:primitive}
k^\alpha \frac{\partial V}{\partial \mathbb{E}}=\mu \tilde{F}_{\rm (rad)}^\alpha.
\end{equation}
Lastly, we split the potential function $V(\boldsymbol{X},\boldsymbol{U})$ in two parts according to (see Eqs. (\ref{eq:splitforce})--(\ref{eq:twoforces2}) and (\ref{eq:potential_def}))
\begin{equation} \label{eq:teopot}
\tilde{F}_{\rm (rad)}(\boldsymbol{X},\boldsymbol{U})^\alpha=-\frac{1}{\mu}\frac{\partial V}{\partial U_\alpha}=-\frac{1}{\mu}\frac{\partial (\mathbb{V}_{\rm C}+\mathbb{V}_{\rm NC})}{\partial U_\alpha},
\end{equation}
where
\begin{eqnarray} 
\mu\mathbb{F}_{\rm C}(\boldsymbol{X},\boldsymbol{U})^\alpha&=&-\frac{\partial \mathbb{V}_{\rm C}}{\partial U_\alpha},\label{eq:teopot21}\\
&&\notag\\
\mu\mathbb{F}_{\rm NC}(\boldsymbol{X},\boldsymbol{U})^\alpha&=&-\frac{\partial \mathbb{V}_{\rm NC}}{\partial U_\alpha}. \label{eq:teopot22}
\end{eqnarray}
We will determine $\mathbb{V}_{\rm C}$ and $\mathbb{V}_{\rm NC}$ in Secs. \ref{sec:PC} and \ref{sec:PNC}, respectively.

\subsubsection{Conservative potential}
\label{sec:PC}
In accordance with our definitions (see Eqs. (\ref{eq:twoforces1}), (\ref{eq:if}), (\ref{eq:primitive}), and (\ref{eq:teopot21})), the potential $\mathbb{V}_{\rm C}$ is defined by
\begin{equation} \label{eq:potC}
-\frac{\partial \mathbb{V}_{\rm C}}{\partial U_\alpha}=-\frac{k^\alpha}{\mathbb{E}}\quad
\Leftrightarrow\quad
-\frac{\partial \mathbb{V}_{\rm C}}{\partial \mathbb{E}}=\frac{1}{\mathbb{E}}.
\end{equation}
In this case, once (\ref{eq:der}) has been exploited, the function $k^\alpha$ simplifies on both members of the second equation in (\ref{eq:potC}), which can be easily integrated in terms of the energy $\mathbb{E}$, yielding
\begin{equation} \label{eq:potC2}
\mathbb{V}_{\rm C}=-\ln(\mathbb{E})+f(\boldsymbol{X},\boldsymbol{U}),
\end{equation}
where $f(\boldsymbol{X},\boldsymbol{U})$ is a function of the local coordinates which is constant with respect to $\mathbb{E}$, i.e.,
\begin{equation} \label{eq:derivata_di_f}
\dfrac{\partial f(\boldsymbol{X},\boldsymbol{U})}{\partial \mathbb{E}}=0.
\end{equation}
To determine $f(\boldsymbol{X},\boldsymbol{U})$, we need to employ the iterative process of integration of exact differential one-forms. Therefore, we have
\begin{equation}
-\frac{\partial \mathbb{V}_{\rm C}}{\partial U_\alpha}=-\frac{k^\alpha}{\mathbb{E}}-\frac{\partial f(\boldsymbol{X},\boldsymbol{U})}{\partial U_\alpha}.
\end{equation}
From the comparison with the corresponding component $\mu \mathbb{F}_{\rm C}(\boldsymbol{X},\boldsymbol{U})^\alpha$, we obtain ${\partial f(\boldsymbol{X},\boldsymbol{U})}/{\partial U_\alpha}=0$.
We have $f(\boldsymbol{X},\boldsymbol{U})=\mbox{const}$, where the constant (\emph{a priori} depending only on the $\boldsymbol{X}$ coordinates) will be determined in Sec. \ref{sec:WFAP}. 

\subsubsection{Non-conservative potential}
\label{sec:PNC}
In this case the joint application of Eqs. (\ref{eq:twoforces2}), (\ref{eq:primitive}), and (\ref{eq:teopot22}) gives for potential $\mathbb{V}_{\rm NC}$
 \begin{equation} \label{eq:newPNC}
-\frac{\partial \mathbb{V}_{\rm NC}}{\partial U_\alpha}=U^\alpha\quad
\Leftrightarrow\quad
k^\alpha \, \frac{\partial \mathbb{V}_{\rm NC}}{\partial \mathbb{E}}=U^\alpha.
\end{equation}
Bearing in mind the condition $U^\alpha U_\alpha=-1$, the second equation in (\ref{eq:newPNC}) can be rearranged as follows: 
\begin{equation} \label{eq:newPNC_2}
\frac{\partial \mathbb{V}_{\rm NC}}{\partial \mathbb{E}} =  \dfrac{1}{\mathbb{E}}.
\end{equation}
Integrating Eq. (\ref{eq:newPNC_2}) with respect to $\mathbb{E}$ leads to
\begin{equation} \label{eq:potPNC2}
\mathbb{V}_{\rm NC}=\ln(\mathbb{E})+f(\boldsymbol{X},\boldsymbol{U}).
\end{equation}
Differentiating Eq. (\ref{eq:potPNC2}) with respect to $U_\alpha$, and using Eq. (\ref{particle_energy_1}), we obtain
\begin{equation}
-\frac{\partial \mathbb{V}_{\rm NC}}{\partial U_\alpha}=\frac{k^\alpha}{(-k^\beta U_\beta)}-\frac{\partial f(\boldsymbol{X},\boldsymbol{U})}{\partial U_\alpha}.
\end{equation}
Comparing the above derivatives with the corresponding components $\mu \mathbb{F}_{\rm NC}(\boldsymbol{X},\boldsymbol{U})^\alpha$ of the radiation force, we find 
\begin{equation} \label{eq:potPNC3}
\begin{aligned}
f(\boldsymbol{X},\boldsymbol{U})&=\int\left[-U^\alpha+\frac{k^\alpha}{(-k^\beta U_\beta)}\right]{\rm d}U_\alpha\\
&=-\int U^\alpha\ {\rm d}U_\alpha-\ln(-k^\beta U_\beta), \;\;\; ({\rm no \; sum \; over}\;  \alpha).
\end{aligned}
\end{equation}
Substituting the last expression in Eq. (\ref{eq:potPNC2}), we obtain $\mathbb{V}_{\rm NC}=-\int U^\alpha\ {\rm d}U_\alpha$. After some algebra, we obtain 
\begin{equation} \label{eq:potPNC4}
\mathbb{V}_{\rm NC}=-\frac{1}{2}U^\alpha U_\alpha +\mbox{const},
\end{equation}
where the integration constant (\emph{a priori} depending only on the $\boldsymbol{X}$ coordinates) will be determined in Sec. \ref{sec:WFAP}.

In conclusion, in the last sections we have realized how our \qm{energy-based strategy} has allowed us to compute in a straightforward way the Rayleigh potential of the PR effect, which reads as
\begin{equation} \label{eq:Rayleigh_potential_final}
V\equiv\mathbb{V}_{\rm C}+\mathbb{V}_{\rm NC} =-\left[ \ln(\mathbb{E})+\frac{1}{2}U_\alpha U^\alpha\right]+\mbox{const}.
\end{equation}

\subsubsection{Classical limit of the Rayleigh potential}
\label{sec:WFAP}
At this stage, it is interesting to check whether the Rayleigh potential (\ref{eq:Rayleigh_potential_final}) is consistent with the classical equations first introduced by Poynting and Robertson \cite{Poynting1903,Robertson1937}. In the classical limit, the components of the radiation force (\ref{eq:radforce1}) become
\begin{eqnarray}
F_{\rm (rad)}(\boldsymbol{X},\boldsymbol{U})^r&\approx&-\frac{A}{r^2}\left(2\dot{r}-1\right), \label{eq:limF1}\\
F_{\rm (rad)}(\boldsymbol{X},\boldsymbol{U})^\varphi&\approx&-\frac{A}{r^2}\left(r\dot{\varphi}\right), \label{eq:limF2}\\
F_{\rm (rad)}(\boldsymbol{X},\boldsymbol{U})^t&=&\dot{r}F_{\rm (rad)}(\boldsymbol{X},\boldsymbol{U})^r+r\dot{\varphi}F_{\rm (rad)}(\boldsymbol{X},\boldsymbol{U})^\varphi
\notag\\
&\approx&-\frac{A}{r^2}\left(-\dot{r}+2\dot{r}^2+r^2\dot{\varphi}^2\right). \label{eq:limF3}
\end{eqnarray}
We remind that factor $\Phi^2$ occurring in Eq. (\ref{eq:radforce1}) can be written as $\Phi^2=\Phi_0^2/r^2$, where $\Phi_0$ is a constant related to the intensity of the radiation field at the emitting surface, and $A=\tilde{\sigma}\Phi^2_0E^2_{\rm p}$ is the luminosity parameter ranging in the interval $[0,1]$, see Refs. \cite{Bini2009,Bini2011,DeFalco20183D,Bakala2019}. Thus, we can relate the radiation force with the derivative of the Rayleigh potential (\ref{eq:Rayleigh_potential_final}) and the integrating factor (\ref{eq:if2}) through (cf. Eq. (\ref{eq:radforce2})) 
\begin{equation} \label{eq:startpoint}
\begin{aligned}
F_{\rm (rad)}(\boldsymbol{X},\boldsymbol{U})^\alpha&=
-\frac{\tilde{\sigma}\Phi_0^2}{r^2}\tilde{F}_{\rm (rad)}(\boldsymbol{X},\boldsymbol{U})^\alpha\\
&=\frac{A}{r^2}\frac{\mathbb{E}^2}{E^2_{\rm p}}\frac{\partial V}{\partial U_\alpha},
\end{aligned}
\end{equation}
where the last equality can be obtained after having multiplied and divided $\tilde{F}_{\rm (rad)}(\boldsymbol{X},\boldsymbol{U})^\alpha$ by $\mu$. The conservative and non-conservative components occurring in the general relativistic Rayleigh potential (\ref{eq:Rayleigh_potential_final}) assume in the classical limit the following form, respectively:
\begin{eqnarray} \label{eq:POTCL1}
\ln(\mathbb{E})&\approx&\ln(E_{\rm p})-\dot{r}-\frac{\dot{r}^2}{2},\label{eq:POTCL1a}\\
\frac{1}{2}U_\alpha U^\alpha&\approx&\frac{1}{2}\left(-1+\dot{r}^2+r^2\dot{\varphi}^2\right). \label{eq:POTCL1b}
\end{eqnarray}
Therefore, in the classical limit we have 
\begin{equation} \label{eq:POTCL2}
V\approx\dot{r}-\frac{1}{2}r^2\dot{\varphi}^2+\left[\frac{1}{2}-\ln (E_{\rm p})\right]+\mbox{const},
\end{equation}
where we can choose 
\begin{equation} \label{eq:constant_term_potential}
\mbox{const}=-\left[\frac{1}{2}-\ln (E_{\rm p})\right],
\end{equation}
to cancel out the term appearing in the square bracket of (\ref{eq:POTCL2}). From Eqs. (\ref{eq:startpoint}) and (\ref{eq:POTCL2}), we have
\begin{equation} \label{eq:startpoint2}
\begin{aligned}
F_{\rm (rad)}(\boldsymbol{X},\boldsymbol{U})^\alpha&\approx\frac{A}{r^2}\left(1-\dot{r}\right)^2\frac{\partial }{\partial U_\alpha}\left(\dot{r}-\frac{1}{2}r^2\dot{\varphi}^2\right),
\end{aligned}
\end{equation}
where Eq. (\ref{eq:Elimit}) has been exploited. Therefore, it is simple to check that Eq. (\ref{eq:startpoint2}) leads immediately to the classical expressions (\ref{eq:limF1}) and (\ref{eq:limF2}) once the underlying derivatives are computed. Since the components $F_{\rm (rad)}(\boldsymbol{X},\boldsymbol{U})^r$ and $F_{\rm (rad)}(\boldsymbol{X},\boldsymbol{U})^\varphi$ deduced from (\ref{eq:startpoint2}) match correctly with their own classical limit, it is obvious that also Eq. (\ref{eq:limF3}) is straightforwardly satisfied. 

\section{Discussion and interpretation of the obtained results}
\label{sec:int_res}
It is extremely important to analyse the results found in the previous sections to focus the attention on several interesting points. Firstly, bearing in mind the factor $-\tilde{\sigma}\Phi^2$ mentioned in Sec. \ref{sec:GRPReffect} and Eqs. (\ref{eq:Rayleigh_potential_final}) and (\ref{eq:constant_term_potential}), the complete analytic expression of the general relativistic Rayleigh potential for the PR effect reads as
\begin{equation} \label{eq:Rayleigh_potential_PR}
V=\tilde{\sigma}\Phi^2\left[\ln\left(\frac{\mathbb{E}}{E_{\rm p}}\right)+\frac{1}{2}\left(U_\alpha U^\alpha+1\right)\right].
\end{equation}
From Ref. \cite{Bini2011,Bakala2019}, we have 
\begin{equation} \label{eq:phi_explicit}
\Phi^2=\Phi^2_0\Bigg/ r^2\left[\frac{\rho}{r^3}\left(1-\frac{2aMb}{\rho}\right)^2-
\left(\frac{b\Delta}{\rho}\right)^2\right]^{1/2},
\end{equation}
where $\Delta=r^2-2Mr+a^2$ and $\rho=r^3+a^2r+2a^2M$, $M$ and $a$ being mass and spin of the black hole (BH), respectively. In addition, remembering that the luminosity parameter is $A=\tilde{\sigma}\Phi^2_0E^2_{\rm p}$, see below Eq. (\ref{eq:radforce2}), and moving
%bring 
$1/E^2_{\rm p}$ outside of the potential, as in Eq. (\ref{eq:startpoint}), we obtain
\begin{equation} \label{eq:potential_explicit}
V=\frac{A}{r^2}\ \frac{\left[\ln\left(\frac{\mathbb{E}}{E_{\rm p}}\right)+\frac{1}{2}\left(U_\alpha U^\alpha+1\right)\right]}{\left[\frac{\rho}{r^3}\left(1-\frac{2aMb}{\rho}\right)^2-\left(\frac{b\Delta}{\rho}\right)^2\right]^{1/2}}.
\end{equation}

After the publication of Poynting's paper in 1903, a fierce controversy arose in the scientific community, which has been originated by the inconsistency of the eponymous effect both with the principles of relativistic mechanics and Maxwell theory of electromagnetism. First attempts to solve such issue were made by J. Larmor and L. Page through aether theory \cite{Larmor1917,Page1918}. The question was partially clarified by H. P. Robertson, who reformulated Poynting's model in the context of special relativity (see Ref. \cite{Robertson1937}, and references therein). Recently, a debate regarding the physical foundations of PR effect between Kla{\v c}ka \emph{et al.} and Burns \emph{et al.} has appeared in the literature \cite{Burns1979,Klacka2014,Burns2014}. 

Einstein theory solves elegantly all delicate difficulties underlying PR model \cite{Bini2009,Bini2011}. In particular, the test particle equations of motion are such that both radiation pressure and PR drag force contributions are included in a single function, i.e., the relativistic radiation force \cite{Klacka2014}. In other words, the general covariance principle prevents any kind of separation between these two terms. On the contrary, such a splitting is admissible only at classical level. The Rayleigh potential reported in Eq. (\ref{eq:potential_explicit}) tremendously support this argument, since it confirms the fact that in a relativistic framework we are not able to distinguish the radiation pressure from the PR effect potential (contrarily to the classical case \cite{Defalco2018}).

Another important feature of the PR model is the dependence of the radiation force on the test particle velocity $\boldsymbol{U}$, see Eq. (\ref{eq:radforce1}). In Sec. \ref{sec:WFAP}, we have noted that in the classical limit the time component $U^t$ is connected with the radiation pressure, while the spatial components, $U^r,U^\theta,U^\varphi$, are linked to the PR drag force. Such a remark implies that the radiation pressure always enters the dynamics, since $U^t$ never vanishes, while the PR drag force could be turned off whenever the test particle is at rest (see Ref. \cite{DeFalco20183D,Bakala2019}, and figures therein for details). It should be noted that this remark does not contradict the general covariance principle, because it is still not possible to separate radiation pressure and PR effect, but it gives insight into the role played by the velocity field in the radiation force.

\subsubsection{Rayleigh potential and test particle trajectory}
\label{sec:RP_TPT}
\begin{figure*}[th!]
\centering
\includegraphics[scale=0.64]{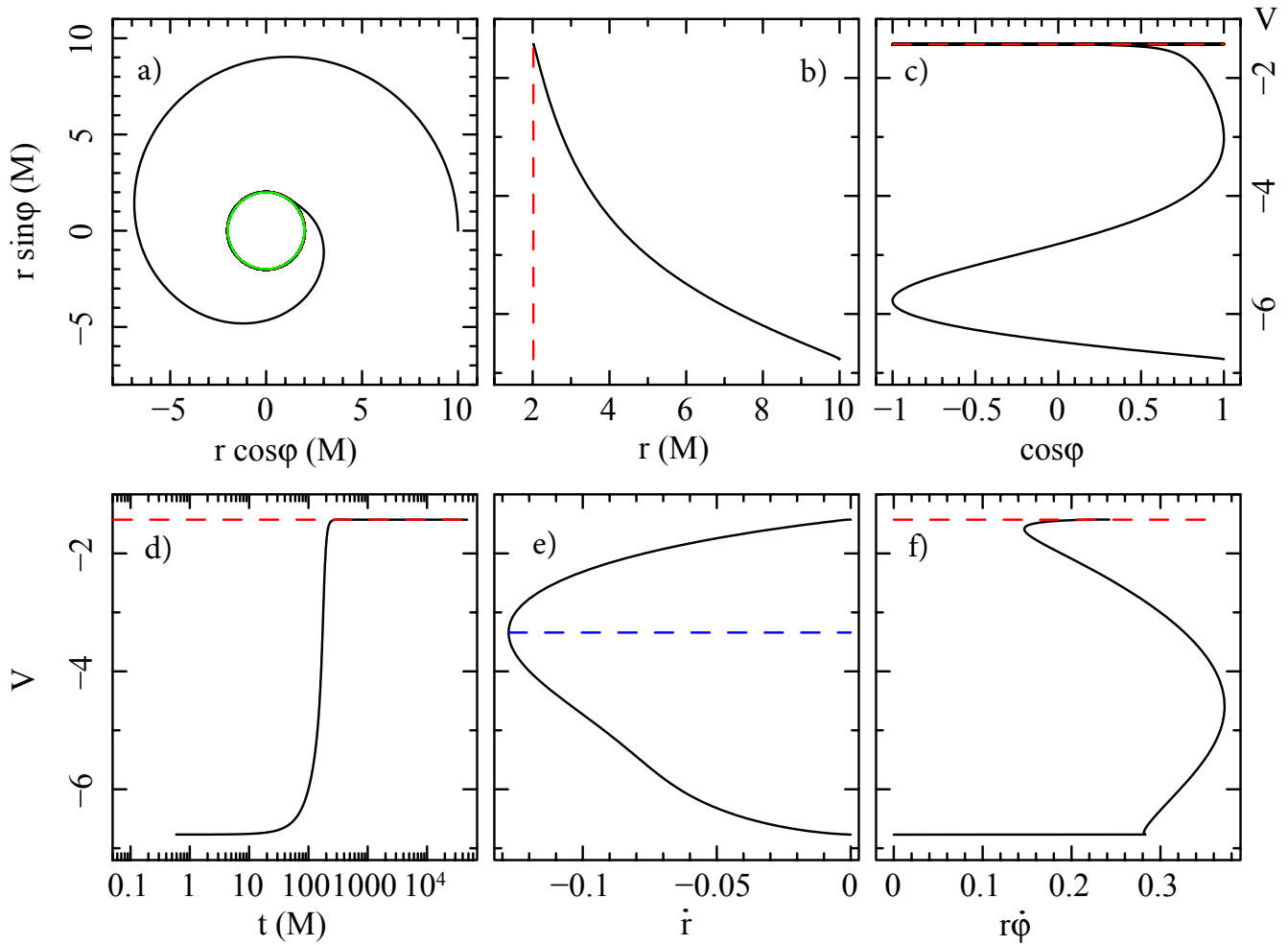}
\caption{Test particle trajectory with the related general relativistic Rayleigh potential (\ref{eq:potential_explicit}) for mass $M=1$ and spin $a=0.1$ of the BH, luminosity parameter $A=0.1$ and photon impact parameter $b=1$. The test particle moves in the spatial equatorial plane with initial position $(r_0,\varphi_0)=(10M,0)$ and velocity $(\nu_0,\alpha_0)=(\sqrt{1/10M},0)$. a) Test particle trajectory spiralling towards the BH and stopping on the critical radius (red dashed line) $r_{\rm (crit)}=2.02M$. The continuous green line is the event horizon radius $r^+_{\rm (EH)}=1.99M$. Rayleigh potential versus b) radial coordinate, c) azimuthal coordinate, d) time coordinate, e) radial velocity, and f) azimuthal velocity. The blue dashed line in panel e) marks the minimum value attained by the radial velocity, corresponding to $\dot{r}=-0.13$. Panels b)--f) must be read from bottom up.}
\label{fig:Fig1}
\end{figure*}

\begin{figure*}[th!]
\centering
\includegraphics[scale=0.64]{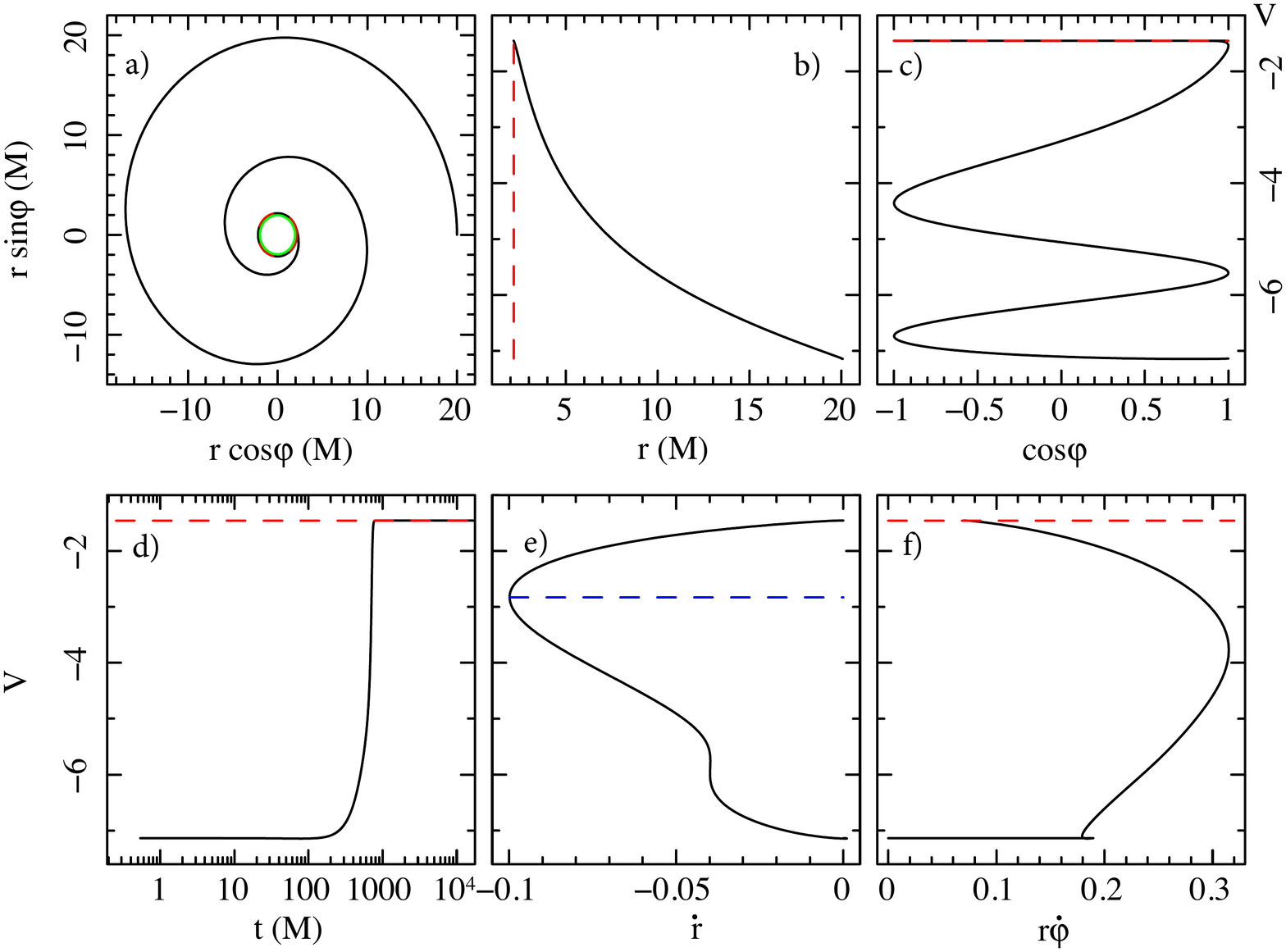}
\caption{Test particle trajectory with the related general relativistic Rayleigh potential (\ref{eq:potential_explicit}) for mass $M=1$ and spin $a=0$ of the BH, luminosity parameter $A=0.3$ and photon impact parameter $b=2$. The test particle moves in the spatial equatorial plane with initial position $(r_0,\varphi_0)=(20M,0)$ and velocity $(\nu_0,\alpha_0)=(0.2,0)$. a) Test particle trajectory spiralling towards the BH and stopping on the critical radius (red dashed line) $r_{\rm (crit)}=2.16M$. The continuous green line is the event horizon radius $r^+_{\rm (EH)}=1.95M$. Rayleigh potential versus b) radial coordinate, c) azimuthal coordinate, d) time coordinate, e) radial velocity, and f) azimuthal velocity. The blue dashed line in panel e) marks the minimum value attained by the radial velocity, corresponding to $\dot{r}=-0.1$. Panels b)--f) must be read from bottom up.} 
\label{fig:Fig2}
\end{figure*}

\begin{figure*}[th!]
\centering
\includegraphics[scale=0.64]{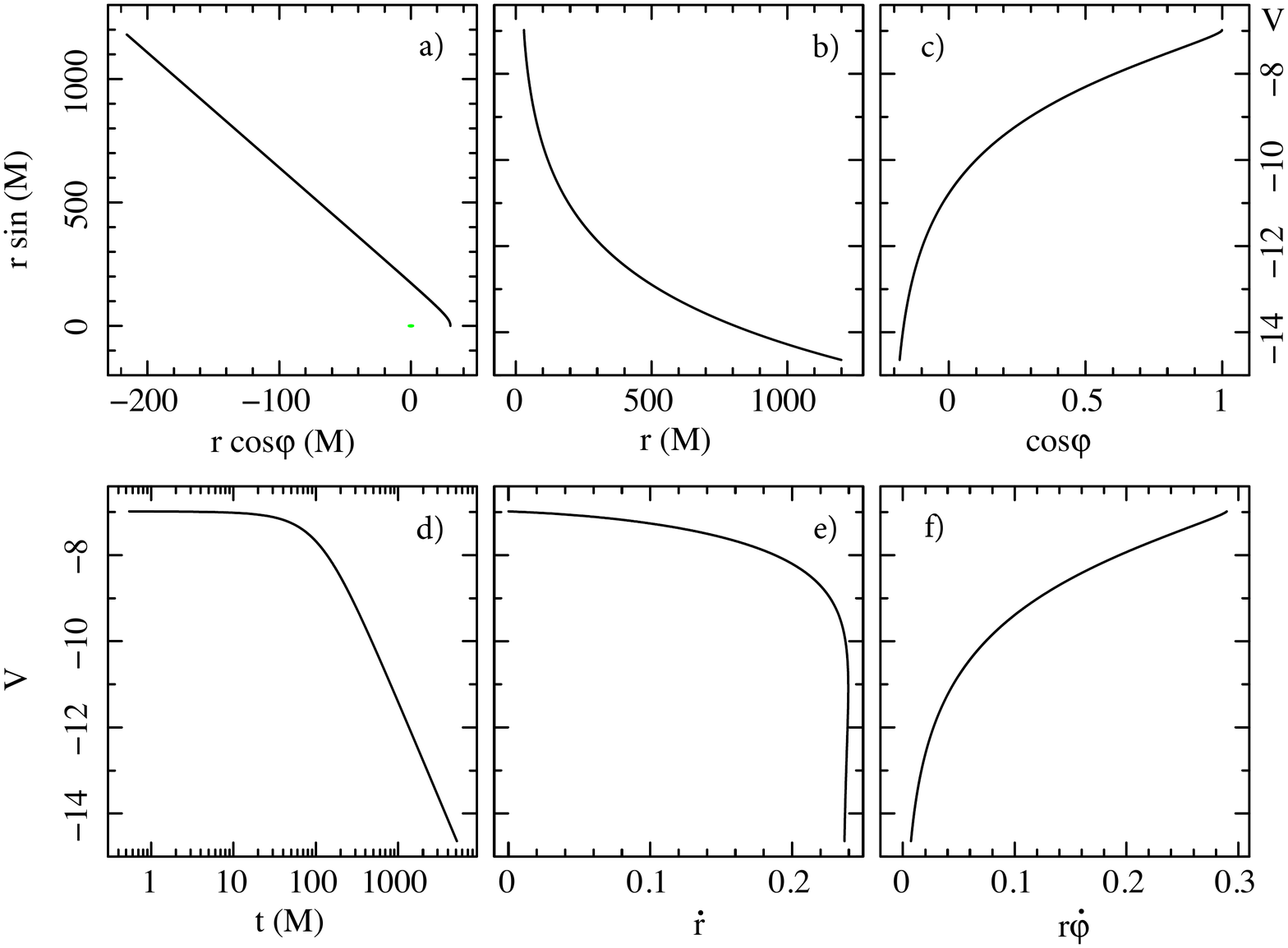}
\caption{Test particle trajectory with the related general relativistic Rayleigh potential (\ref{eq:potential_explicit}) for mass $M=1$ and spin $a=0.8$ of the BH, luminosity parameter $A=0.8$ and photon impact parameter $b=5$. The test particle moves in the spatial equatorial plane with initial position $(r_0,\varphi_0)=(30M,0)$ and velocity $(\nu_0,\alpha_0)=(0.3,0)$. a) Test particle trajectory departing from the BH and approaching spatial infinity. The continuous green line is the event horizon radius $r^+_{\rm (EH)}=1.6M$. Rayleigh potential versus b) radial coordinate, c) azimuthal coordinate, d) time coordinate, e) radial velocity, and f) azimuthal velocity. Panels b)--f) must be read from top down.} 
\label{fig:Fig3}
\end{figure*}

We recall that the test particle can end its motion either at infinity or on the critical hypersurface (region where there is a balance between radiation and gravitational forces) \cite{Bini2009,Bini2011,DeFalco20183D,Bakala2019}. In the Schwarzschild case, the test particle can either stop on a point of the critical hypersurfuce (for $b=0$), or move on it with constant velocity, equal to the azimuthal photon velocity (for $b\neq0$); on the other hand, in the Kerr case the test particle will always move with constant velocity on the critical hypersurface due to the frame dragging effect \cite{Bini2009,Bini2011,DeFalco20183D,Bakala2019}.

The analytical form of the Rayleigh potential is relevant for its strong correspondence with the test particle trajectory, creating thus a direct link with observations. 

From Fig. \ref{fig:Fig1}, it is possible to note explicitly this connection. Indeed in panel a), we see that the test particle spirals inward around a slowly rotating BH (in Kerr metric) of mass $M=1$ and spin $a=0.1$, having a luminosity parameter $A=0.1$ with a photon impact parameter $b=1$. The test particle motion ends on the critical radius $r_{\rm crit}=2.02M$ (dashed red line), very close to the event horizon $r^+_{\rm (EH)}\equiv1+\sqrt{1-a^2}=1.99M$ (continuous green line), where it starts corotating with constant velocity around the BH, due to the frame dragging effect and the radiation field (see \cite{Bini2009,Bini2011,DeFalco20183D,Bakala2019}, for details). 

To gain further information on the test body dynamics and the involved radiation processes, we analyse the Rayleigh potential in terms of different variables. Panels b)--f) must be read from bottom up. In panel b), we note that the Rayleigh potential increases almost exponentially with respect to the radial coordinate $r$, starting from the position $r_0=10M$ until the final destination $r=r_{\rm (crit)}$. In panel c), we analyse the Rayleigh potential through the azimuthal coordinate $\varphi$. The initial azimuthal position is $\cos\varphi_0=1$ (i.e., $\varphi_0=0$). Counting how many times the potential profile comes back to the initial position, we can deduce the number of windings, $n_{\rm wind}$, around the BH. In our example, $n_{\rm wind}$ amounts to $1$, as can be easily checked from panel a). The motion ends when the potential reaches its maximum (red dashed line), where the test particle moves with constant velocity on the critical region without changing the value of its Rayleigh potential. 

In Fig. \ref{fig:Fig1}d, we study the behaviour of the Rayleigh potential with respect to the time coordinate $t$. The profile assumes a distinctive $S$-shape, passing from the initial minimum value $V\sim-6.77$ to its maximum $V\sim-1.43$ with a jump in time of $t_{\rm jump}\sim800M$. In order to help the reader figure out how small the latter value is, we calculate four time estimations regarding some relevant astrophysical situations: for the smallest and lightest stellar BH ever observed, knwon as XTE J1650--500 (having mass $M=3.8M_\odot$ \cite{Shaposhnikov2009}), we see that the jump lasts $t_{\rm jump}\sim15.02$ ms; instead for the heaviest stellar BH, GW150914 (with mass $M=62M_\odot$ \cite{Abbott2016}), we obtain $t_{\rm jump}\sim0.25$ s; for an intermediate BH, like the the one recently discovered at the center of 47 Tucanae (having a mass of $M=2300M_\odot$ \cite{Kiziltan2017}), the jump results to be $t_{\rm jump}\sim9.09$ s; finally, for a supermassive BH, like SgrA* in the center of our own Galaxy (having a mass of $M=4\times10^6M_\odot$ \cite{Gillessen2017}), we evaluate a $t_{\rm jump}\sim4.39$ h. 

In panel e), we analyse the Rayleigh potential in terms of the radial velocity, i.e., $\dot{r}\equiv dr/dt$. The graph, possessing vanishing radial velocity both at its starting-point and end-point, assumes a quasi-parabolic shape and has the dashed blue line at $V\sim-3.34$ as quasi-symmetric axis. This profile expresses the fact that the test particle starts decelerating until it reaches the value $\dot{r}\sim-0.13$ (occurring, as can be inferred from plots b) and c), at the position $(r,\varphi)\approx(3M,0.28)$), where it accelerates before smoothly braking on the critical region.  Last panel f) shows the behaviour of the Rayleigh potential with respect to the azimuthal velocity $r\dot{\varphi}\equiv rd\varphi/dt$. The azimuthal velocity is initially zero, afterward it gets its maximum value $r\dot{\varphi}\sim0.37$, then it tends to decrease, and finally to increase again (due to the frame dragging effect), until it attains the constant value $r\dot{\varphi}\sim0.24$.   

In Fig. \ref{fig:Fig2}a we plot the motion of a test particle orbiting a static BH of mass $M=1$ (described in the Schwarzschild metric), affected by luminosity $A=0.3$ with photon impact parameter $b=2$. Like before, the test particle spiral motion ends on the critical region $r_{\rm crit}=2.16M$ (red dashed line). Panels b)--f) must be read from bottom up. We note the following similarities with the former case: in panel b) we can appreciate the typical quasi-exponential grow of the $V$ potential; plot c) expresses the winding of the test particle ($n_{\rm wind}=2$ in this example); panel d) returns again the characteristic $S$-shape, where the jump occurs at $t_{\rm jump}\sim700M$; plot e) exhibits the maximum deceleration line for the radial velocity (dashed blue line) at $\dot{r}=-0.1$; graph f) demonstrates how the test particle acquires the maximum azimuthal velocity $r\dot{\varphi}\sim0.32$ before drifting down to the critical radius without increasing its velocity, because there is no frame dragging effect.

Figure \ref{fig:Fig3} refers to the last case analysed. It deviates considerably from the previous two. Indeed in Fig. \ref{fig:Fig3}a, the test particle moves around a rotating BH of mass $M=1$ and spin $a=0.8$ (extreme regime in Kerr metric), endowed with an intense luminosity $A=0.8$ and a photon impact parameter $b=5$. Dynamical motion terminates at spatial infinity, i.e., no critical radius appears. The initial amount of energy suffices to let the test body escape from the two combined attracting forces, i.e., the gravitational pull and the PR drag force. However, the PR effect still continues to remove energy from the test particle, which nevertheless has enough energy (velocity) to be not dragged towards the critical hypersurface. In this particular case, the high-value of the luminosity parameter $A$ offers, through the radiation pressure, a great contribution to the test particle for escaping to infinity.  In this case, the Rayleigh potential exhibits new features, not encountered before (panels b)--f) must be read from top down). Indeed, in panel b), the Rayleigh potential decreases quasi-exponentially, while in c) it shows a decreasing-monotone behaviour, without winding up around the BH, i.e., $n_{\rm wind}=0$. In graph d), we end up with a reversed trend, since an almost linear decay-shape arises. Unlike the former cases, the potential begins with a maximum value $V\sim-6.76$ and start decreasing at time $t_{\rm dec}\sim60M$ toward the minimum value $V\sim-14.21$. In panel e), we learn that the test particle increases velocity, due to the weakening of both the gravitational pull and the PR drag force. However, asymptotically it should approach the value $\dot{r}\sim0.23$. Also the azimuthal velocity graph f) differentiates itself from Figs. \ref{fig:Fig1}f and \ref{fig:Fig2}f. Indeed, the test particle initially has $r\dot{\varphi}\sim0.3$, then it slows down until it is asymptotically at the rest. 

The huge difference between Figs. \ref{fig:Fig1} and \ref{fig:Fig2} (i.e., test body approaching the critical region) and Fig. \ref{fig:Fig3} (i.e., test particle going to infinity) relies mainly on the three following evidences: ($i$) the Rayleigh potential profile is negative and changes, assuming peculiar and recognizable features for both the situations; ($ii$) if the test particles spirals inward, the Rayleigh potential is monotone-increasing, otherwise it is monotone-decreasing; ($iii$) depending on the case, as a consequence of the remark ($ii$), the plots b)--f) must be read either from bottom up (spiraling inward) or from up down (getting to infinity).             

We decided to draw the test body dynamics in the equatorial plane only \cite{Bini2009,Bini2011}, because the three-dimensional model gives exactly the same results, except that two more plots, related to the $\theta$-motion (position and velocity), must be considered in this case \cite{DeFalco20183D,Bakala2019}. In addition, we have considered for simplicity in all figures $\alpha_0=0$, because for $\alpha_0\neq0$ there are not great differences. Regarding the module of the initial velocity $\nu_0$, it is important to take into account the Keplerian velocity $\nu_K$. Its expression can be obtained by setting to zero the expression for $d\alpha/d\tau$ in the PR equation of motion (see Eq. (2.34) in Ref. \cite{Bini2011}), and considering $\alpha=0$ ($\alpha=\pi$) for co-rotating (counter-rotating) motion. In this way, we end up with the following equation for $\nu_K$:

\begin{equation}
\begin{aligned}
&a(n)^{\hat r}+2\nu_K \theta(n)^{\hat r}{}_{\hat \varphi}+k_{\rm (Lie)}(n)^{\hat r}\nu_K^2\\
&=A\frac{(1+bN^\varphi)(1-\nu_K\cos\beta)}{\gamma N^2(g_{\theta\theta}g_{\varphi\varphi})^{1/2}},
\end{aligned}
\end{equation}
which is a function of the initial position $r_0$, the spin $a$, the luminosity parameter $A$, and the photon impact parameter $b$. The Keplerian velocity derives from the balance of the gravitational part with the radiation effects. We immediately note that for $A=0$, we obtain the classical definition in pure gravity. Based on the above definition, we think that this velocity plays a fundamental role in determing the motion of the test particle. Indeed, if the initial velocity is Keplerian or sub-Keplerian, i.e., $\nu_0\le\nu_K$, the test particle spirals down to the critical hypersurface, otherwise for super-Keplerian initial velocity, $\nu_0>\nu_K$, the test particle has enough energy to escape to infinity. Therefore, the Keplerian velocity $\nu_K$ can be considered as the threshold between confined and unbounded motions. A fundamental role is also played by the luminosity parameter $A$, where for high luminosities, $A\gtrsim0.75$, we have noted that the radiation pressure dominates over the PR effect imparting thus supplementary energy and angular momentum to the test particle to spiral outward. Therefore, if a test particle moves on a circular orbit with Keplerian velocity, it posses already half the energy needed to reach infinity, and the remaining half could be provided by the radiation force, if this dominates over gravity and PR effect. However, the three displayed examples have only the purpose of showing how the Rayleigh potential changes for confined (see Figs. \ref{fig:Fig1} and \ref{fig:Fig2}) and unbounded (see Fig. \ref{fig:Fig3}) motions. In this paper we would like just to present the potentiality of our formalism in general astrophysical contexts, and leave the deep investigation of the implications between dynamics and Rayleigh potential to another paper.

The plots presented here assign an enormous value to the Rayleigh potential for its observational features. Indeed, by monitoring the test bodies motion around a rotating/static BH it will be possible to infer useful proprieties on the involved radiation processes and, in particular, to reconstruct the functional form of the Rayleigh potential by means of observational techniques through plots b)--f). This step will in turn allow to derive analytically the related radiation force and deduce crucial proprieties regarding the gravitational field (e.g., BH mass and spin), radiation processes (e.g., radiation intensity and photon impact parameter), and dissipation effects (e.g., the role played by the PR effect and the attitude of the system to be influenced by it).

Viceversa, our analysis can also be employed for theoretical purposes. Indeed, by assigning a different functional form to the Rayleigh potential one can straightforwardly calculate the test particle equations of motion and the related trajectories. In such a way, the theoretical investigation of a wide range of radiation processes, including the possible dissipative phenomena, becomes an easy task. Finally, our model allows to benchmark theoretical and synthetic results with observational data.

\subsubsection{Digression on logarithmic Rayleigh potential}
\label{sec:LOGPOT}
The Rayleigh potential (\ref{eq:potential_explicit}) contains the \emph{logarithmic} term $\ln(\mathbb{E}/E_{\rm p})$. This represents a novel aspect in the literature involving relativistic dissipation in radiation processes. In the framework of potential theory, such function has been adopted in different research fields. The implications of a logarithmic potential in Schr\"odinger \cite{Birula1976} and Klein-Gordon equations \cite{Rosen1969,Bartkowski2008} have been examined in the context of non-relativistic quantum mechanics and quantum field theory, respectively. In addition, this kind of nonlinearity appears naturally in inflation cosmology \cite{Barrow1995}, galactic dynamics models \cite{Valluri2012}, and supersymmetric field theories \cite{Enqvist1998}. Besides, there have been profound developments in polynomial and rational approximation theory \cite{Saff2010}, whereas in measure theory it has been fundamental to solve problems arising in electrostatic and classical gravity \cite{Saff1997}. Finally, a recent application consisted in approximating (through a logarithm) the gravitational potential in the regions close to a Schwarzschild BH to analytically describe the motion of test particles and accretion disk structures \cite{Shakura2018}. 

In our model, the logarithmic function is quenched far from the BH by the factor $1/r^2$, while close to it the logarithm dominates, see Eq. (\ref{eq:potential_explicit}). From Sec. \ref{sec:PC}, we realize that the term $\ln(\mathbb{E}/E_{\rm p})$ can be ascribed to the conservative part of the radiation force, $\mathbb{F}_{\rm C}{}^\alpha$. In the classical limit, we obtain
\begin{equation} \label{eq:FCCL}
\mathbb{F}_{\rm C}{}^\alpha\approx\frac{A}{r^2}(1-\dot{r})[1,1,0,0],
\end{equation}
where we recognize that the azimuthal force is zero, i.e., $\mathbb{F}_{\rm C}{}^\varphi=0$. Instead, the radial force is given by 
\begin{equation} \label{eq:FCCL_rad}
\mathbb{F}_{\rm C}{}^r\approx\frac{A}{r^2}-\frac{A}{r^2}\dot{r},
\end{equation}
where $A/r^2$ represents the radiation pressure and $A\dot{r}/r^2$ is half the radial PR drag force, see Eq. (\ref{eq:limF1}). Since in the classical limit the radiation field is constituted by photons travelling along straight lines, radiation absorption occurs only in radial direction. The test particle energy is less than the incoming photon energy, i.e., $\mathbb{E}\le E_{\rm p}$ (Figs. \ref{fig:Fig1} -- \ref{fig:Fig3} confirms that $\ln(\mathbb{E}/E_{\rm p})$ is everywhere negative). Consequently, Eq. (\ref{particle_energy_1}) configures as an absorption energy describing the interaction between the test particle and the radiation field. Indeed, when the test particle is at rest, $\mathbb{E}$ reaches its maximum value (i.e., $\mathbb{E}=E_{\rm p}$), reflecting the fact that the photon energy is entirely absorbed, whereas as the test particle velocity approaches the speed of light, $\mathbb{E}$ tends to zero, since in this case the photon can not hit the test body. In other words, the faster the test body-target moves, the more photon-bullets are \qm{dissipated}, since the absorbed energy strongly depends on the test particle velocity. More formally, when the test particle moves at the speed of light ($\nu\to1$), we have $\mathbb{E}\to0$, and therefore the radiation force (\ref{eq:splitforce}) tends to zero. Therefore, we conclude that the logarithmic function is associated with the \emph{relativistic absorption processes}. 

Bearing in mind the previous observations, we can figure out that the term $U_\alpha U^\alpha$ appearing in Eq. (\ref{eq:potential_explicit}) and stemming from the non-conservative part $\mathbb{F}_{\rm NC}{}^\alpha$ of the radiation force (see Sec. \ref{sec:PNC}), describes the \emph{re-emission process} in the radial and azimuthal directions. The condition according to which the norm of the test particle four-velocity is constrained to be $-1$, reflects the fact that re-emission is constant, isotropic, and independent of $\boldsymbol{U}$. In addition, since the time component $U^t$ is always nonzero, re-emission will always be present (at least until absorption occurs). 

All the absorbed radiation is afterward completely re-emitted by the test particle, which thus behaves as an ideal black body in thermal equilibrium. Indeed, in our model, absorption and re-emission are intimately intertwined for two reasons. Firstly, the test particle four-velocity $U^\alpha$ appears both in $\mathbb{V}_{\rm C}$ and $\mathbb{V}_{\rm NC}$. Moreover, both mechanisms would quit if $U^\alpha$ could be replaced with $k^\alpha$, i.e., if the test particle velocity gets really close to the speed of light. 

The above results agree with the hypotheses according to which both absorption and re-emission configure as PR effect causes. Furthermore, it should be clear that, although it is not possible to separate covariantly radiation pressure from PR drag contributions (as stated in Sec. \ref{sec:int_res}), we can clearly distinguish absorption from re-emission moments anyway.

\section{Conclusions}
\label{sec:end}
Interaction with the environment represents the main feature which differentiates dissipative systems from conservative ones. The investigation of the former represents a demanding task, as briefly outlined in the Introduction. 

In this paper, we have dealt with a specific inverse problem in the calculus of variations involving dissipation. In particular, via the combined use of an integrating factor (which has proved to be crucial to make the differential semi-basic one-form (\ref{PR_semibasic_oneform}) exact) and an integration strategy built on the energy dissipated by the system, we have obtained, for the first time in the literature, the analytical form of the Rayleigh potential of the general relativistic PR effect, see Sec. \ref{sec:AEF}. 
Equations (\ref{eq:splitforce}) and (\ref{eq:der}) constitute the core of our \qm{energy-based strategy}, guaranteeing a twofold advantage. Firstly, a sensible reduction of the calculations, underlying the determination of the $V$ potential related to the exact differential semi-basic one-form $\boldsymbol{\omega}$, is achieved, since the integration process involves only the energy $\mathbb{E}$, instead of the four variables $\boldsymbol{U}$. In addition, the obtained expression of the $V$ potential, as a function of $\mathbb{E}$, suffices for the description of the dynamics. This might represent a crucial point in all those situations in which, unlike the case discussed in our paper, the evaluation of $f(\boldsymbol{X},\boldsymbol{U})$ turns out to be too laborious, see Secs. \ref{sec:PC} and \ref{sec:PNC}.

Furthermore, we have shown how the Rayleigh potential permits to recover all the known features of the PR effect, see Sec. \ref{sec:int_res}. In addition, in agreement with the spirit of the Lagrangian approach, the main information of the dynamical system under study is enclosed in one single function. We have discovered new interesting implications of the general relativistic Rayleigh potential, which can be summarised in the following points:

\begin{itemize}
 \item it is not possible to split radiation pressure and PR effect in the GR frame. These two forces are merged in one single function, in agreement with the general covariant principle, see Sec. \ref{sec:int_res}. Moreover, the radiation pressure is always present and it is linked to the time component of the velocity field, while the PR effect can be turned off for particular configurations at rest, having therefore a close connection with the spatial components of the velocity field; 
    
\item the most significant contributions are displayed in Figs. \ref{fig:Fig1}--\ref{fig:Fig3}, where we have described how to link coherently the observations with the theoretical results and viceversa. This allows to acquire useful information on the mathematical structure and on the physical properties of the phenomenon analysed. These results might confer a prominent observational relevance to the PR model;
    
\item a new functional class, represented by the logarithm function occurring in Eq. (\ref{eq:potential_explicit}), has been discovered. At the best of our knowledge, this is a new facet in the literature. We have ascertained that such term describes absorption processes. This may entail significant implications in the context of both PR effect and, more in general, radiation processes in high-energy astrophysics. In particular, we have interpreted  $\mathbb{E}$ as the dissipated energy (cf. Eq. (\ref{particle_energy_1})) since, by considering the photons as the bullets shot against test particle (acting as the target), it gives information about the photons being \qm{wasted} due to the test body motion;
    
\item the term $U_\alpha U^\alpha$ in Eq. (\ref{eq:potential_explicit}) is related to the radiation re-emission properties and embodies the hypothesis according to which the test particle behaves as an ideal black body in thermal equilibrium. In addition, absorption and re-emission processes, despite being intimately connected through their dependence on the velocity field $\boldsymbol{U}$, can be separated without violating the relativistic covariant principle (see Sec. \ref{sec:LOGPOT});  
   
\item the Rayleigh potential (\ref{eq:potential_explicit}) includes also gravitational effects. Apart from the $\Phi$ term (cf. Eq. (\ref{eq:phi_explicit})), notably these are included in the velocity field $\boldsymbol{U}$ and the dissipated energy $\mathbb{E}$. This reveals \emph{a fortiori} the (general) relativistic character of the Rayleigh potential.   
    
\end{itemize}
 
 As pointed out before, the aim of our strategy consists in obtaining the analytic form of the Rayleigh potential through the use of the dissipated energy variable, since it entails a drastic reduction of the underlying calculations. Therefore, a natural question arises: \emph{might it be possible to generalise such integration strategy to solve the inverse problem of calculus of variations related to dissipative systems in GR?}.  This is a very  delicate and  interesting issue, which, we deem, can be answered in two complementary ways: 

\begin{itemize}
\item[(1)] \emph{general-inductive approach}: one should characterise the functional class of the exact (naturally closed in a simply connected domain) differential semi-basic one-forms. The use of an integrating factor is an advantageous mechanism enlarging the set of dissipative systems admitting closed one-forms. However, this path is fairly challenging, because the mathematical information on such functional space is minimal, being limited  to the closure condition only. However, a more encouraging approach relies on examining  particular sets of functions, which may probably be useful to mine the whole functional space. 

\item[(2)] \emph{particular-deductive approach}: one should find other examples, besides the general relativistic PR effect, which can be analytically solved in terms of Rayleigh potential function. This route might involve some astrophysical models discussed in the literature. Such specific examples can encompass different mathematical functions and hence, along with the first approach, can give more insights into the mathematical structure of dissipative systems. Indeed, as part of our future research program, we intend to exploit such integration strategy in the  context of gravitational waves. 

\end{itemize}

\section*{Acknowledgements}
The authors are grateful to Professor Luigi Stella, Doctor Giampiero Esposito, Professor Giuseppe Marmo, and Professor Tom Mestdag for the stimulating discussions and the useful suggestions aimed at improving the scientific impact of this formalism. The authors are grateful to Professor Antonio Romano for the useful discussions. The authors thank the Silesian University in Opava and the International Space Science Institute in Bern for hospitality and support. The authors are grateful to Gruppo Nazionale di Fisica Matematica of Istituto Nazionale di Alta Matematica for support. The authors thank the anonymous referee for the valuable comments aimed at improving the physical aspects of the paper.
\bibliography{references}
\end{document}